\documentclass{PoS}
\usepackage{cite}
\def\g{\gamma}
\def\m{\mu^2}
\def\G{\Gamma}
\def\i{\prime}
\def\am{(\alpha_1-\beta_1)}
\def\ap{(\alpha_1+\beta_1)}
\def\amn{\am_{\pi^0}}
\def\amc{\am_{\pi^{\pm}}}
\def\apc{\ap_{\pi^{\pm}}}
\def\apn{\ap_{\pi^0}}
\def\bm{(\alpha_2-\beta_2)}
\def\bp{(\alpha_2+\beta_2)}
\def\bmn{\bm_{\pi^0}}
\def\bmc{\bm_{\pi^{\pm}}}
\def\bpc{\bp_{\pi^{\pm}}}
\def\bpn{\bp_{\pi^0}}
\def\gg{\g \g\to \pi^0 \pi^0}
\def\s{\sigma}

\def\ggc{\g\g\to\pi^+\pi^-}
\def\gp{\g p\to\g\pi^+n}
\def\pia{\pi^-A\to\g\pi^-A}

\def\mpp{M_{++}}
\def\mm{M_{+-}}
\def\sp{s'}
\def\tp{t'}
\def\unit{10^{-4} {\rm fm}^3}
\def\unitq{10^{-4} {\rm fm}^5}
\def\be{\begin{equation}}
\def\ee{\end{equation}}
\def\beq{\begin{eqnarray}}
\def\eeq{\end{eqnarray}}

\title{Dispersion sum rules for pion polarizabilities}

\ShortTitle{Dispersion sum rules for pion polarizabilities}

\author{\speaker{Lev Fil'kov}%
        \thanks{Financial support by the research center "EMG" is
        gratefully acknowledged.}\\
        Lebedev Physical Institute\\
        E-mail: \email{filkov@sci.lebedev.ru}}

\author{Viktor Kashevarov\\
        Lebedev Physical Institute\\
        E-mail: \email{kashev@kph.uni-mainz.de}}

\abstract{An analysis of different dispersion sum rules (DSRs) for the dipole 
          and quadrupole pion polarizabilities is carried out.
          We prove the absence of additional spurious singularities in these 
          approaches. The results of the calculations of the polarizabilities
          in the framework of DSRs are compared with ChPT predictions and with 
          data obtained in different experiments.}

\FullConference{6th International Workshop on Chiral Dynamics, CD09\\
		July 6-10, 2009\\
		Bern, Switzerland}

\begin{document}

\section{Introduction}

Pion polarizabilities are fundamental structure parameters characterizing
the behavior of the pion in an external electromagnetic field.
The dipole electric ($\overline{\alpha}_1$) and magnetic 
($\overline{\beta}_1$) pion polarizabilities
measure the response of the pion to quasistatic electric and magnetic
fields. On the other hand, the quadrupole polarizabilities 
$\overline{\alpha}_2$ and $\overline{\beta}_2$
measure the electric and magnetic quadrupole moments induced in the pion
in the presence of an applied field gradient.

The generalized dipole ($\alpha_1$ and $\beta_1$)
and quadrupole ($\alpha_2$ and $\beta_2$) polarizabilities are defined 
\cite{rad,fil2} through
expansion of the non-Born helicity amplitudes of Compton scattering on 
the pion in powers of $t$ at fixed $s=\m$
\beq
\mpp(s=\m,t)&=&\pi\mu\left[2\am+\frac{t}{6}\bm\right]+{\cal O}(t^2),
\nonumber \\
\mm(s=\m,t)&=&\frac{\pi}{\mu}\left[2\ap+\frac{t}{6}\bp\right]+{\cal O}(t^2),
\label{mpm}
\eeq
where $s=(q_1+k_1)^2$, $t=(k_1-k_2)^2$ ($q_1,\;q_2$ and $k_1,\;k_2$ are the
pion and photon four-momenta), and $\mu$ is the pion mass.
In the following the dipole and quadrupole polarizabilities are given in
units $\unit$ and $\unitq$, respectively.

It should be noted that the generalized pion polarizabilities very
strongly differ from intrinsic polarizabilities $\overline{\alpha}_i$ 
and $\overline{\beta}_i$ \cite{petr1}.
For example, the nonrelativistic expression of
the generalized electric dipole polarizability $\alpha_1$ 
is equal to
\be
\alpha_1=\overline{\alpha}_1+\frac{\alpha}{3\mu}<r_{\pi}^2>=
\overline{\alpha}_1 + 15.5,
\ee
where $<r_{\pi}^2>$ is the mean-square pion radius, 
$\alpha$ is the fine-structure constant.

The expression for the generalized magnetic dipole polarizability $\beta_1$ is
more complicated. Additional contributions could be both paramagnetic and
diamagnetic. So, it is difficult to determine the nature of magnetic
susceptibility of $\beta_1$ at present.

In the following we will omit the word "generalized".

The values of the pion polarizabilities are very sensitive to predictions of
different theoretical models. Therefore, accurate experimental determination
of these parameters is very important for testing the validity of such
models.

For example, the results of calculations of $\amc$
in Refs \cite{fil2,fil3,rad2,fil1} are at variance with the predictions of
chiral perturbation theory (ChPT) \cite{gasser2,burgi}. On the other hand, 
P. Pasquini, D. Drechsel, and S. Scherer (PDS) in Ref. \cite{pds} 
claim that as the absorptive part of the Compton amplitudes 
in Refs \cite{fil2,fil3,rad2,fil1} is expressed by Breit-Wigner poles 
with energy dependent coupling 
constants and decay widths, there must appear additional
spurious singularities. As a result, the values of the polarizabilities 
obtained in \cite{fil2,fil3,rad2,fil1} have to be modified essentially.

In the present paper we examine the statement of PDS and give an overview 
of the present situation in the field of investigation of the dipole and 
quadrupole polarizabilities of the charged and neutral pions in the
frameworks of different DSRs and ChPT and compare the results of calculations
with available experimental data.

\section{Dispersion sum rules for dipole polarizabilities}

DSRs for the difference and sum of electric and magnetic pion polarizabilities
have been constructed using dispersion relations (DRs) for the helicity
amplitudes $\mpp$ and $\mm$, respectively. It has been shown in 
Ref. \cite{aberb} that these amplitudes have no kinematical singularities 
or zeroes.

In Refs \cite{fil2,fil3,fil1} DSR has been constructed for  
$\am$ using DR for the amplitude $\mpp$ at fixed $u=\m$ (where $u=2\m-s-t$).
In this case, the Regge-pole model allows the use of DR without subtractions 
\cite{aberb}. Such a DSR is
\be
\am=\frac{1}{2\pi^2\mu}\left(\int\limits_{4\m}^{\infty}~\frac{
Im\mpp(\tp,u=\m)~d\tp}{\tp}
+\int\limits_{4\m}^{\infty}~\frac{Im\mpp(\sp,u=\m)~d\sp}{\sp-\m}\right).
\label{dsr}
\ee
The imaginary parts of the amplitudes in these DR and DSR are approximated 
by the contributions of meson resonances using Breit-Wigner expressions. 
For example, the contributions of the vector mesons are determined as
\be
Im\mpp^{(V)}(s,t)=-4g_V^2 s\frac{\G_0}{(M_V^2-s)^2+\G_0^2}  
\ee
\label{im}
for $s>4\m$ and $Im\mpp^{(V)}(s,t)=0$ for $s<4\m$, where 
\be
g_V^2=6\pi\sqrt{\frac{M_V^2}{s}}\left(\frac{M_V}{M_V^2-\m}\right)^3
\G_{V\to\g\pi}, \qquad
\G_0=\left(\frac{s-4\m}{M_V^2-4\m}\right)^{\frac32}M_V\G_V.
\ee
Here $M_V$, $\G_V$, and $\G_{V\to \g\pi}$ are the mass, full and $\g\pi$ decay 
widths of the vector mesons, respectively. A dependence of the width on 
the energy is conditioned by the threshold behavior. The energy dependence of 
the coupling constant $g_V$ 
appears via an expression for the total cross section of the process
$\g\pi\to\g\pi$ through the vector meson contribution. 

In order to check the possibility of the appearance of additional singularities
in our dispersion approach, 
we calculate the contributions of all mesons, except $\s$, 
to our DSR by the zero-width approximation.

The results of such calculations of $\am_{z}$ 
are listed in Tables 1, 2 together with the complete calculations of 
$\am_{f}$ obtained in Ref. \cite{fil2}.
\begin{table}[ht]
\caption{The DSR predictions for $\amc$.}
\centering
\begin{tabular}{ccccccccccc}\hline
 &$\rho$&$b_1$&$a_1$&$a_2$&$f_0$&$f_0^{\i}$&$\sigma$&$\Sigma$
&$\Delta\Sigma$ \\ \hline
$\am_{f}$&-1.15&0.93 &2.26 &1.51 &0.58&0.02&9.45&13.60&2.15 \\ \hline
$\am_{z} $&-1.11&0.85 &3.39 &1.51 &0.59&0.03&9.45&13.70&     \\ \hline
\end{tabular}
\end{table}
\begin{table}[ht]
\caption{The DSR predictions for $\amn$.}
\centering
\begin{tabular}{cccccccccc}\hline
 &$\rho$&$\omega$&$\phi$&$f_0$&$f_0^{\i}$&$\sigma$&$\Sigma$
&$\Delta\Sigma$ \\ \hline
$\am_{f}$&-1.58 &-12.56&-0.04 &0.60 &0.02 &10.07&-3.49  &2.13 \\ \hline
$\am_{z} $&-1.99 &-11.81&-0.04 &0.61 &0.02 &10.07&-3.14  &     \\ \hline
\end{tabular}
\end{table}

As seen from the Tables the zero-width 
approximation results practically coincide with the calculations 
of Ref. \cite{fil2} which are beyond such an approximation. 

The coefficient $1/\sqrt{t}$ in the coupling constant of the $\sigma$-meson 
amplitude in Ref.\cite{fil2,fil3} provides the correct asymptotic behavior 
for the convergence of the 
integral over $t$ in our DSR and does not lead to additional singularities. 
To check it we have calculated this integral using the energy independent 
values of the decay width and the coupling constant of the $\sigma$-meson.
It has not lead to essential changes of the calculation results presented in
the Tables. For example, the value of $\am_{\pi^{\pm}z}$ would be equal to 
13.1.

Besides, we compare our DSR calculation results in the $s$-channel with
the predictions of DSR obtained at the fixed angle 
$\theta_{\gamma\pi}=180^{\circ}$ \cite{rad2}. In this case  
\be
\am^{(s)}=\frac{1}{2\pi^2}\int\limits_{3\mu/2}^{\infty}~\frac{d\nu}{\nu^2}
\left[ 1+\frac{\nu}{\mu}\right] [\sigma(yes)-\sigma(no)],
\ee
\label{angl}
where $\nu$ is the incident photon energy in the lab. system,
$\sigma(yes)$ and $\sigma(no)$ stand for the sum of the photoabsorption cross 
sections containing, respectively, parity-flip and -nonflip multipoles. 

The best way to calculate this integral would be the use of experimental values 
for the cross sections. On the other hand, if these experimental data are well 
described by some function in the physical region of the process under 
consideration, then we can use this function, but only in this region, without 
continuation into unphysical regions. Usual Breit-Wigner forms
for the photoabsorption cross sections
with energy dependent decay widths and correct asymptotic behavior 
are such functions. Therefore, we should use them only
in the physical regions of these processes. 
If we add a contribution from spurious singularities,
which are out of the physical region considered in DSR, we would have 
an additional contribution to the result obtained from the integration of
the experimental cross section. This is a gross mistake.
So, there are no problems 
with additional spurious singularities in the derivation and the calculation 
of Eq. (2.4). The results of the calculation of this 
expression are very close to the values of $\am_{f}$ given in Tables (1) and 
(2) for $\rho$, $\omega$, $\phi$, $a_1$, and $a_2$ mesons which 
saturate the DSR integrals in the $s$-channel. This result confirms the absence 
of additional singularities in our approach for the $s$-channel integral of our 
DSRs. The $t$-channel contributions with $I=J=0$ are the same for both DSR at
fixed $u=\m$ and DSR at fixed $\theta_{\gamma\pi}=180^{\circ}$.
There are the same arguments why additional singularities are absent in 
the $t$-channel too.

It should be noted that in the work of S.S. Kamalov, L. Tiator, 
D. Drechsel {\em et al.} \cite{kam}  
the neutral pion photoproduction and electroproduction at the threshold 
were analyzed using DR. In this work 
the resonance contribution to the imaginary parts of the amplitudes of the 
pion photoproduction and electroproduction were given in terms of Breit-Wigner
expressions with the energy dependent decay widths and coupling constants. 
According to the main statement of PDS, 
there are very many additional singularities in the resonance amplitudes 
considered in this work.
The results of the calculations in the work \cite{kam} were obtained without
consideration of any additional singularities and are
in very good agreement with the experimental data for the 
pion photoproduction in the threshold region.  
However, if, according to PDS, one takes into account these singularities, 
this would lead to additional contributions and, as a result, to
a disagreement with the experiment. 
This confirms that an account of such singularities is a mistake.

It is worth noting that the calculation of $\amc$ in the framework of DSR   
at finite energy \cite{petr2}, which takes into account the $s$-channel 
and the Regge-pole asymptotic contribution only, yielded $\amc=10.3\pm 1.3$. 
This value practically coincides with our result (see Table 1). 

As for the sum $\ap$, these values are calculated using Baldin's
DSR
\be
\ap=\frac{1}{2\pi^2}~\int\limits_{\frac32 \mu}^{\infty}~\frac{
\s_T(\nu) d\nu}{\nu^2},
\ee
\label{dsr1p}
where $\s_T$ is the total cross section of the $\g\pi$-interaction.
These DSR results in
\beq
&& \apc=0.166\pm 0.024, \nonumber \\ 
&& \apn=0.802\pm 0.035.
\eeq

On the other hand, two-loop ChPT calculations \cite{gasser2,gasser1}
give
\beq
&& \amc=5.7\pm 1.0,    \nonumber \\
&& \apc=0.16,          \nonumber \\
&& \amn=-1.9\pm 0.2,   \nonumber \\
&& \apn=1.1\pm 0.3 .
\eeq

So, the results of the ChPT calculations for
the sum and difference of the dipole polarizabilities of $\pi^0$ and
the sum for the charged pions do not conflict within the errors with
predictions of DSRs.

Let us consider possible reasons of the discrepancy between the 
predictions of DSRs and ChPT for $\amc$.
The main contribution to the DSRs for $\amc$ is given by the $\sigma$-meson.
However, this meson is taken into account only partially
in the present ChPT calculations.

Consider the methods of the calculation of the vector meson
contribution in the frameworks of DSRs and ChPT. 
In the narrow width approximation we have from Eq. (2.2)
$$
Im M^{(V)}_{++}(s,t)=
-\frac{4}{\pi} g_V^2 s \delta (s-M_v^2).
$$
Then the DSR calculation gives 
\be\label{vdsr}
Re \mpp(s=\m,t=0)=\frac{-4g_V^2M_V^2}{(M_v^2-\m)}.
\ee
In the case of ChPT the authors of Ref. \cite{gasser2} used
\be\label{vchpt}
Re M_{++}(s=\m,t=0)=\frac{-4g_V^2\m}{(M_V^2-\m)}.
\ee
The absolute value of the amplitude (2.9) is smaller than (2.8) by
a factor $M_V^2/\m$. From the point of view of analyticity, the result (2.9) 
could be obtained if DR with one subtraction at $s=0$ is used for the 
amplitude $\mpp(s,t)$. However, an additional subtraction constant 
$\mpp(s=0,t=0)$ then appears, 
which was not considered in the available ChPT calculations.

In the case of the difference of the dipole polarizabilities of the 
$\pi^0$-meson,
the big contribution of the $\sigma$-meson to DSR is cancelled by the big 
contribution
of the $\omega$-meson. On the other hand, in the ChPT calculations the 
$\sigma$-meson is only partially 
included and the $\omega$-meson gives a very small 
contribution to this difference. As a result, the DSR and ChPT predictions
for $\amn$ are rather close.

\section{Experimental data for dipole polarizabilities of charged pions}

By now the values of the pion polarizabilities were determined by analyzing
the processes $ \pi^-A\to \g\pi^-A$, $\gp$, and $\g\g\to\pi\pi$.
The experimental information available so far for the difference of the
dipole polarizabilities of charged pions is summarized in Table 3.

\begin{table}
\caption{The experimental data presently available for $\amc$. 
In \cite{serp,comp,bab,don,kal}
$\amc$ was determined by using the constraint
$\alpha_{1\pi^\pm}=-\beta_{1\pi^\pm}$.}
\centering
\begin{tabular}{|ll|l|} \hline
\multicolumn{2}{|c|}{Experiments} & \qquad\quad $\amc$  \\ \hline
$\gp$ & MAMI (2005) \cite{mami} 
& $11.6\pm 1.5_{stat}\pm 3.0_{syst}\pm 0.5_{mod}$ \\ \hline
$\gp$ & Lebedev Phys. Inst. (1984) \cite{lebed}& $40\pm 20$ \\ \hline
$\pia$ & Serpukhov (1983) \cite{serp} & $13.6\pm 2.8\pm 2.4$ \\ \hline  
$\pia$ & COMPASS (2007) \cite{comp} & $5.0\pm 3.4$ \quad (preliminary)
\\ \hline\hline
\multicolumn{2}{|c|}
{$\ggc$ \quad ($E_{\g}< 700$ MeV)} &    \\ \hline
\multicolumn{2}{|c|}
{D. Babusci {\em et al.} (1992) \cite{bab}} &   \\
     & PLUTO \cite{pluto}   & $38.2\pm 9.6\pm 11.4$ \\
     & DM 1 \cite{dm1}     & $34.4\pm 9.2$  \\
     & DM 2 \cite{dm2}     & $52.6\pm 14.8$  \\
     & MARK II \cite{mark} & $4.4\pm 3.2$    \\ \hline
\multicolumn{2}{|c|}
{J.F. Donoghue, B.R. Holstein (1993) \cite{don}} &  5.4 \\
\multicolumn{2}{|c|}  
{MARK II \cite{mark}}  &       \\ \hline
\multicolumn{2}{|c|}  
{A.E. Kaloshin, V.V. Serebryakov (1994) \cite{kal}} & $5.25\pm 0.95$ \\
\multicolumn{2}{|c|}
{MARK II \cite{mark}}   &         \\ \hline\hline
\multicolumn{2}{|c|}
{L.V. Fil'kov, V.L. Kashevarov (2006) \cite{fil3}}&  $13^{+2.6}_{-1.9}$ \\
$\ggc$ & fit of data \cite{mark,tpc,cello,ven,aleph,belle}&   \\
       & from threshold to 2.5 GeV  &  \\ \hline
\end{tabular}
\end{table}

The values of the experimental cross sections of the process $\ggc$ in the
energy region $E_{\g}<700$ MeV are very ambiguous. As a result, the values
of $\amc$, obtained from analyses of these data, lie in the interval
4.4--52.6. The analyses of the data of Mark II \cite{mark} only
have given $\amc$ close to the ChPT result. 

The difference $\amc$ found from the global fit to all available
experimental data of the process $\ggc$  
in the energy region from the threshold to 2500 MeV \cite{fil2} agrees 
very well with the results \cite{serp} obtained from the scattering of 
high energy $\pi^-$ mesons off the Coulomb field of heavy nuclei 
and from the radiative photoproduction of $\pi^+$ from the proton at MAMI
\cite{mami} and in Lebedev Physical Institute \cite{lebed} (see Table~3)
and with the DSR calculations.
However, these values of $\amc$ deviate essentially
from the ChPT calculations \cite{gasser2,burgi}.

Results of polarizability extraction from the process 
$\pi^-A\to \gamma\pi^-A$ strongly depend on the
momentum transfer $Q^2$. In this reaction the Coulomb amplitude
dominates for $Q^2\lesssim 10^{-4}$ (GeV/c)$^2$. In the region of 
$Q^2\sim 10^{-3}$ (GeV/c)$^2$ Coulomb and nuclear contributions are of 
similar size. In this region the nuclear contribution, in particular
an interference between the Coulomb and nuclear amplitudes, should be taken
into account. In the work \cite{serp} the authors considered 
$Q^2<6\times 10^{-4}$ (Gev/c)$^2$, while the authors of Ref.\cite{comp}
worked at $Q^2<7.5\times 10^{-3}$ (Gev/c)$^2$. In this region of $Q^2$
the contribution of the interference between the Coulomb and nuclear 
amplitudes is very large \cite{fil4,walch}. However, it was not
taken into account in the work \cite{comp}. 
This is the main reason of the difference between the Serpukhov and
COMPASS results. Moreover, for the
total energy, in the $\gamma\pi$ c.m.s., $W\gtrsim 450$ MeV 
and $\theta_{\gamma\gamma^{\prime}}\sim 180^{\circ}$, the $\sigma$-meson 
contribution should be taken into account \cite{fil1}.

\section{Pion quadrupole polarizabilities}

DSRs for the difference and the sum of the quadrupole polarizabilities
have been obtained with the help of DRs at fixed $u=\m$ with one
subtraction for the amplitudes $\mpp$  and $\mm$, respectively:
\be
\bm=\frac{6}{\pi^2\mu}\left\{\int\limits_{4\m}^{\infty}~\frac{
Im\mpp(\tp,u=\m)~d\tp}{\tp^2} -\int\limits_{4\m}^{\infty}~\frac{
Im\mpp(\sp,u=\m)~d\sp}{(\sp-\m)^2}\right\},
\label{dsrm}
\ee
\be
\bp=\frac{6\mu}{\pi^2}\left\{\int\limits_{4\m}^{\infty}~\frac{
Im\mm(\tp,u=\m)~d\tp}{\tp^2}-\int\limits_{4\m}^{\infty}~\frac{
Im\mm(\sp,u=\m)~d\sp}{(\sp-\m)^2}\right\}.
\label{dsrp}
\ee

The corresponding values of the quadrupole polarizabilities have 
been found in Refs \cite{fil2,fil3}
by fitting the experimental total cross sections of the 
processes $\gg$ \cite{mars,bien} and $\ggc$ 
\cite{mark,tpc,cello,ven,aleph,belle} 
in the energy regions from the thresholds 
to 2.25 and 2.5 GeV, respectively. The fit functions were constructed
by using DRs with subtractions for the amplitudes $\mpp$ and $\mm$.  

The values of the quadrupole polarizabilities found in Refs. 
\cite{fil2,fil3} and the predictions of DSRs \cite{fil2} and 
ChPT \cite{gasser2,gasser1} are listed in Table~4.
\begin{table}
\caption{The quadrupole polarizabilities of the neutral and charged pions.}
\centering
\begin{tabular}{|c|l|l|c|c|} \hline
         &                       &                &\multicolumn{2}{|c|}{
ChPT \cite{gasser2,gasser1}} \\ \cline{4-5}
     &\qquad fit \cite{fil2,fil3} &DSR \cite{fil2} &to one-loop & to two-loops
\\ \hline
$\bmn$   &$39.7\pm0.02$        &$39.72\pm 8.01$  &    &$37.6\pm 3.3$ \\ \hline
$\bpn$   &$-0.181\pm 0.004$    &$-0.171\pm 0.067$&    & 0.04        \\ \hline
$\bmc$   &$25.0^{+0.8}_{-0.3}$ &$25.75\pm 7.03$  &11.9&16.2 [21.6]  \\ \hline
$\bpc$   &$0.133\pm 0.015$     &$0.121\pm 0.064$ &  0 &-0.001 [-0.001] \\ \hline
\end{tabular}
\end{table}
The numbers in brackets correspond to the order $p^6$ low energy constants
from Ref. \cite{nc}.
As seen from this Table, all values of the polarizabilities found in Ref.
\cite{fil2, fil3} are in good agreement with the DSR predictions \cite{fil2}.

The difference of the quadrupole polarizabilities $\bmc$ obtained in 
Refs.\cite{fil2,fil3} 
disagrees with the present two-loop ChPT calculations \cite{gasser2,gasser1}.
One of the sources of this disagreement is the bad knowledge of the low energy
constants. Moreover, it should be noted that in this case the two-loop
contribution generates nearly 100\% as compared to the one-loop result.
The ChPT calculations of $(\alpha_2+\beta_2)$ give an opposite sign. However,
calculations of $(\alpha_2+\beta_2)$ at order $p^6$ determine only the
leading order term in ChPT. Therefore, contributions at $p^8$ could be
essential, and considerably more work is required to put the chiral 
prediction on a firm basis in this case \cite{gasser2}. 

It is worth noting that calculations of the dipole and quadrupole pion 
polarizabilities in the frame of the Nambu-Jona-Lasino model \cite{hill}
agree within errors with the DSR \cite{fil2,fil3} predictions.
 
\section{Summary}

We showed that there are no problems with additional spurious singularities
in the DSRs and DRs considered. The difference between the predictions of the 
DSRs and ChPT for $\amc$ remains. This discrepancy is 
connected with a different account of the contribution of the $\sigma$ and 
vector mesons in the DSR and ChPT calculations. The disagreement between 
the DSR and ChPT predictions of the quadrupole polarizabilities is connected, 
in particular, with the bad knowledge of the low energy constants. 
Substantial corrections to the values of the quadrupole polarizabilities 
are expected from three-loop calculations. 
\vspace{0.1cm}
                                                             
The authors thank A.I. L'vov and V.A. Petrun'kin for useful discussions.
This research was supported by the DFG-RFBR (Grant No. 09-02-91330).

\end{document}